\documentclass[]{spie}
\usepackage[]{graphicx}                                                                  

\newcommand{\ud}{\mathrm{d}}

\title{Localization and Pattern Formation in Quantum Physics. \\
II. Waveletons in Quantum Ensembles}

\author{Antonina ~N. Fedorova and Michael ~G. Zeitlin
\skiplinehalf
\supit{}  IPME RAS, St.~Petersburg, V.O. Bolshoj pr., 61, 199178, Russia
}
\authorinfo{http://www.ipme.ru/zeitlin.html,
http://www.ipme.nw.ru/zeitlin.html,
E-mail: zeitlin@math.ipme.ru, anton@math.ipme.ru}   

\begin{document}

\begin{center}                                                                       
\begin{tabular}{p{160mm}}                                                            
                                                                                     
\begin{center}                                                                       
{\bf\Large                                                                           
LOCALIZATION AND PATTERN FORMATION} \\                                                    
\vspace{5mm}                                                                         
                                                                                     
{\bf\Large IN QUANTUM PHYSICS.}\\                                           
\vspace{5mm}                                                                         
                                                                                     
{\bf\Large II. WAVELETONS IN QUANTUM ENSEMBLES}\\                                                        
                                                                                     
\vspace{1cm}                                                                         
                                                                                     
{\bf\Large Antonina N. Fedorova, Michael G. Zeitlin}\\                               
                                                                                     
\vspace{1cm}

{\bf\large\it                                                                        
IPME RAS, St.~Petersburg,                                                            
V.O. Bolshoj pr., 61, 199178, Russia}\\                                              
{\bf\large\it e-mail: zeitlin@math.ipme.ru}\\                                        
{\bf\large\it e-mail: anton@math.ipme.ru}\\                                          
{\bf\large\it http://www.ipme.ru/zeitlin.html}\\                                     
{\bf\large\it http://www.ipme.nw.ru/zeitlin.html}                                    
                                                                                   
\end{center}                                                                         
                                                                                     
\vspace{1cm}                                                                         
                                                                                     
\begin{abstract}  
In this second part we present a set of methods, analytical and
numerical, which can describe behaviour in 
(non) equilibrium ensembles, both classical and quantum, especially in
the complex systems, where the standard approaches cannot
be applied. The key points demonstrating advantages of this approach are:
(i) effects of localization of
possible quantum states; (ii) effects of non-perturbative multiscales 
which cannot be calculated by means of perturbation
approaches; (iii) effects of formation of complex/collective 
quantum patterns from
localized modes and classification and possible control of the full zoo of
quantum states, including (meta) stable localized patterns (waveletons).
We demonstrate the appearance of nontrivial localized (meta) stable
states/patterns in a number of collective models covered by the 
(quantum)/(master) hierarchy 
of Wigner-von Neumann-Moyal-Lindblad equations, 
which are the result of ``wignerization'' procedure (Weyl-Wigner-Moyal quantization)
of classical BBGKY kinetic hierarchy, and present the
explicit constructions for exact analytical/numerical computations. Our
fast and efficient approach is based on variational and multiresolution
representations in the bases of polynomial tensor algebras of generalized
localized states (fast convergent variational-wavelet
representation). We construct the representations for hierarchy/algebra of
observables(symbols)/distribution functions via the complete multiscale
decompositions, which allow to consider the polynomial and rational type
of nonlinearities. The solutions are represented via the exact
decomposition in nonlinear high-localized eigenmodes, which correspond to
the full multiresolution expansion in all underlying hidden time/space or
phase space scales. In contrast with different approaches we do not use
perturbation technique or linearization procedures. 
Numerical modeling shows the creation of
different internal structures from localized modes, which are related to
the localized (meta) stable patterns (waveletons), entangled ensembles
(with subsequent decoherence) and/or chaotic-like type of behaviour. 
\end{abstract} 
\vspace{5mm}

\begin{center}
{\large Submitted to Proc. of SPIE Meeting,
The Nature of Light: What is a Photon?\\
Optics \& Photonics, SP200, }
{\large San Diego, CA, July-August, 2005}

\vspace{5mm}

\end{center}                                                                         
\end{tabular}                                                                        
\end{center}                
\newpage

\maketitle                                             

\begin{abstract}  

In this second part we present a set of methods, analytical and
numerical, which can describe behaviour in 
(non) equilibrium ensembles, both classical and quantum, especially in
the complex systems, where the standard approaches cannot
be applied. The key points demonstrating advantages of this approach are:
(i) effects of localization of
possible quantum states; (ii) effects of non-perturbative multiscales 
which cannot be calculated by means of perturbation
approaches; (iii) effects of formation of complex/collective 
quantum patterns from
localized modes and classification and possible control of the full zoo of
quantum states, including (meta) stable localized patterns (waveletons).
We demonstrate the appearance of nontrivial localized (meta) stable
states/patterns in a number of collective models covered by the 
(quantum)/(master) hierarchy 
of Wigner-von Neumann-Moyal-Lindblad equations, 
which are the result of ``wignerization'' procedure (Weyl-Wigner-Moyal quantization)
of classical BBGKY kinetic hierarchy, and present the
explicit constructions for exact analytical/numerical computations. Our
fast and efficient approach is based on variational and multiresolution
representations in the bases of polynomial tensor algebras of generalized
localized states (fast convergent variational-wavelet
representation). We construct the representations for hierarchy/algebra of
observables(symbols)/distribution functions via the complete multiscale
decompositions, which allow to consider the polynomial and rational type
of nonlinearities. The solutions are represented via the exact
decomposition in nonlinear high-localized eigenmodes, which correspond to
the full multiresolution expansion in all underlying hidden time/space or
phase space scales. In contrast with different approaches we do not use
perturbation technique or linearization procedures. 
Numerical modeling shows the creation of
different internal structures from localized modes, which are related to
the localized (meta) stable patterns (waveletons), entangled ensembles
(with subsequent decoherence) and/or chaotic-like type of behaviour 
\end{abstract} 
  

\section{INTRODUCTION: CLASSICAL AND QUANTUM ENSEMBLES}

In this paper we consider the  applications of 
a nu\-me\-ri\-cal\--\-ana\-ly\-ti\-cal technique based on local nonlinear harmonic analysis
to  the description of quantum ensembles.
The corresponding class of individual Hamiltonians has the form
\begin{eqnarray}
\hat{H}(\hat{p},\hat{q})=\frac{\hat{p}^2}{2m}+U(\hat{p},\hat{q}),
\end{eqnarray}
where $U(\hat{p}, \hat{q})$ is an arbitrary polynomial 
function on $\hat{p}$, $\hat{q}$, 
and plays the key role in many areas of physics [1]. Many cases, 
related to some physics models, are considered in [2]-[8].
It is a continuation of our more qualitative approach considered in part I [9].
In this part our goals are some attempt 
of classification and the constructions of explicit numerical-analytical representations 
for the existing quantum states in the class of models described (non) equilibrium ensembles
and related collective models. 
There is a hope on the understanding of relation between the structure of initial Hamiltonians and
the possible types of quantum states and the qualitative type of their behaviour.
Inside the full spectrum  there are at least three possibilities which are the most 
important from our point of view:
localized states, chaotic-like or/and entangled patterns, localized (stable) patterns 
(qualitative definitions/descriptions can be found in part I [9]).
All such states are interesting in the different areas of physics [1]. 
Our starting point is the general point of view of a deformation quantization approach at least on
the naive Moyal/Weyl/Wigner level.
The main point of such approach is based on ideas from [1], which allow  
to consider the algebras of quantum observables as the deformations
of commutative algebras of classical observables (functions). 
So, if we have  the classical counterpart of 
Hamiltonian (1) as a model for classical dynamics 
and the Poisson manifold $M$ (or symplectic 
manifold or Lie coalgebra, etc) as the corresponding phase space, 
then for quantum calculations we need first of all to find
an associative (but non-commutative) star product 
 $*$ on the space of formal power series in $\hbar$ with
coefficients in the space of smooth functions on $M$ such that
\begin{eqnarray}
f * g =fg+\hbar\{f,g\}+\sum_{n\ge 2}\hbar^n B_n(f,g), 
\end{eqnarray}
where
$\{f,g\}$
is the Poisson brackets, $B_n$ are bidifferential operators.
In the naive calculations we may use the simple formal rule:
\begin{eqnarray}
* &\equiv&\exp \Big(\frac{i\hbar}{2}(\overleftarrow\partial_q\overrightarrow\partial_p-
   \overleftarrow\partial_p\overrightarrow\partial_q)\Big)
\end{eqnarray}

In this paper we consider the calculations of the Wigner functions
$W(p,q,t)$ (WF) corresponding
to the classical polynomial Hamiltonian $H(p,q,t)$ as the solution
of the Wigner-von Neumann equation [1]:
\begin{eqnarray}
i\hbar\frac{\partial}{\partial t}W = H * W - W * H
\end{eqnarray}
and related Wigner-like equations for different ensembles.
According to the Weyl transform, a quantum state (wave function or density 
operator $\rho$) corresponds
to the Wigner function, which is the analogue in some 
sense of classical phase-space distribution [1].
We consider the following operator 
form of differential equations for time-dependent WF, $W=W(p,q,t)$:
\begin{eqnarray}
W_t=\frac{2}{\hbar}\sin\Big[\frac{\hbar}{2}
(\partial^H_q\partial^W_p-\partial^H_p\partial^W_q)\Big]\cdot HW
\end{eqnarray}
which is a result of the Weyl transform 
or ``wignerization'' of von Neumann equation for density matrix:
\begin{equation}
i\hbar\frac{\partial\rho}{\partial t}=[H,\rho]
\end{equation}

\subsection{BBGKY Ensembles}

We start from  set-up for kinetic BBGKY hierarchy 
(as c-counterpart of proper q-hierarchy).
We present the explicit analytical construction for solutions of
both hierarchies of equations, which is based 
on tensor algebra extensions of multiresolution
representation and variational formulation.
We give explicit representation for hierarchy of n-particle 
reduced distribution functions 
in the base of
high-localized generalized coherent (regarding underlying generic symmetry 
(affine group in the simplest case)) 
states given by polynomial tensor algebra of our basis functions 
(wavelet families), which 
takes into account
contributions from all underlying hidden multiscales
from the coarsest scale of resolution to the finest one to
provide full information about stochastic dynamical process.
The difference between classical and quantum case is concetrated
in the structure of the set of operators included in the set-up and, surely,
depends on the method of quantization.
But, in the naive Wigner-Weyl approach for quantum case the symbols of operators  
play the same role as usual functions in classical case. 
In some sense,
our approach resembles Bogolyubov's one and related approaches 
but we don't use any perturbation technique (like virial expansion)
or linearization procedures.
Most important, that 
numerical modeling in both cases shows the creation of
different internal (coherent)
structures from localized modes, which are related to stable (equilibrium) or 
unstable type
of behaviour and
corresponding pattern (waveletons) formation.

Let M be the phase space of ensemble of N particles ($ {\rm dim}M=6N$)
with coordinates
\begin{eqnarray}
x_i=(q_i,p_i), \quad i=1,...,N,\qquad
q_i=(q^1_i,q^2_i,q^3_i)\in R^3,\qquad
p_i=(p^1_i,p^2_i,p^3_i)\in R^3,\qquad
q=(q_1,\dots,q_N)\in R^{3N}.
\end{eqnarray}
Individual and collective measures are: 
\begin{eqnarray}
\mu_i=\ud x_i=\ud q_i\ud p_i,\quad \mu=\prod^N_{i=1}\mu_i
\end{eqnarray}
Distribution function
$D_N(x_1,\dots,x_N;t)$
satisfies 
Liouville equation of motion for ensemble with Hamiltonian $H_N$ and normalization constraint:
\begin{eqnarray}
\frac{\partial D_N}{\partial t}=\{H_N,D_N\},\qquad 
\int D_N(x_1,\dots,x_N;t)\ud\mu=1
\end{eqnarray}
where Poisson brackets are:
\begin{eqnarray}
\{H_N,D_N\}=\sum^N_{i=1}\Big(\frac{\partial H_N}{\partial q_i}
\frac{\partial D_N}{\partial p_i} - \frac{\partial H_N}{\partial p_i}
\frac{\partial D_N}{\partial q_i}\Big)
\end{eqnarray}
Our constructions can be applied to the following general Hamiltonians:
\begin{eqnarray}
H_N=\sum^N_{i=1}\Big(\frac{p^2_i}{2m}+U_i(q)\Big)+
\sum_{1\leq i\leq j\leq N}U_{ij}(q_i,q_j)  
\end{eqnarray}
where potentials 
$U_i(q)=U_i(q_1,\dots,q_N)$ and $U_{ij}(q_i,q_j)$
are not more than rational functions on coordinates.
Let $L_s$ and $L_{ij}$ be the Liouvillean operators (vector fields)
\begin{eqnarray}
L_s=\sum^s_{j=1}\Big(\frac{p_j}{m}\frac{\partial}{\partial q_j}-
\frac{\partial u_j}{\partial q}\frac{\partial}{\partial p_j}\Big)-\sum_{1\leq i
\leq j\leq s}L_{ij},\qquad
L_{ij}=\frac{\partial U_{ij}}{\partial q_i}\frac{\partial}{\partial p_i}+
\frac{\partial U_{ij}}{\partial q_j}\frac{\partial}{\partial p_j}
\end{eqnarray}
For s=N we have the following representation for Liouvillean vector field
$
L_N=\{H_N,\cdot \}
$
and the corresponding ensemble equation of motion:
\begin{eqnarray}
\frac{\partial D_N}{\partial t}+L_ND_N=0
\end{eqnarray}
$L_N$ is 
self-adjoint operator regarding standard pairing on the set of phase space 
functions.
Let
\begin{eqnarray}
F_N(x_1,\dots,x_N;t)=\sum_{S_N}D_N(x_1,\dots,x_N;t)
\end{eqnarray}
be the N-particle distribution function ($S_N$ is permutation group of N element
s). 
Then we have the hierarchy of reduced distribution functions ($V^s$ is the
corresponding normalized volume factor) 
\begin{eqnarray}
F_s(x_1,\dots,x_s;t)=
V^s\int D_N(x_1,\dots,x_N;t)\prod_{s+1\leq i\leq N}\mu_i
\end{eqnarray}
After standard manipulations we arrived to c-BBGKY hierarchy:
\begin{eqnarray}
\frac{\partial F_s}{\partial t}+L_sF_s=\frac{1}{\upsilon}\int\ud\mu_{s+1}
\sum^s_{i=1}L_{i,s+1}F_{s+1}
\end{eqnarray}
It should be noted that we may apply our approach even to more general formulation. 
As in the general as in particular situations (cut-off, e.g.) 
we are interested in the cases when
\begin{eqnarray}
F_k(x_1,\dots,x_k;t)=\prod^k_{i=1}F_1(x_i;t)+G_k(x_1,\dots,x_k;t),
\end{eqnarray}
where $G_k$ are correlators, really have additional reductions 
as in the simplest case of one-particle truncation 
(Vlasov/Boltzmann-like systems). So, the proper dynamical formulation is reduced to the
(infinite) set of equations for correlators/partition functions.
Then by using physical motivated reductions or/and during 
the corresponding cut-off procedure 
we obtain, 
instead of linear and pseudodifferential (in general case)
equations,
their finite-dimensional but nonlinear approximations with
the polynomial type of nonlinearities (more exactly, multilinearities).
Our key point in the following consideration is the proper 
generalization of naive perturbative multiscale Bogolyubov's structure
restricted by the set of additional physical hypotheses.

\subsection{Quantum ensembles}

Let us start from the second quantized representation for an algebra of observables

\begin{equation}
A=(A_0,A_1,\dots,A_s,...)
\end{equation}
in the standard form

\begin{eqnarray}
A=A_0+\int dx_1\Psi^+(x_1)A_1\Psi(x_1)+\dots+
(s!)^{-1}\int dx_1\dots dx_s\Psi^+(x_1)\dots
\Psi^+(x_s)A_s\Psi(x_s)\dots
\Psi(x_1)+\dots
\end{eqnarray}
N-particle Wigner functions

\begin{eqnarray}
W_s(x_1,\dots,x_s)&=&\int dk_1\dots dk_s{\rm exp}\big(-i\sum^s_{i=1}k_ip_i\big)
{\rm Tr}\rho\Psi^+\big(q_1-\frac{1}{2}\hbar k_1\big)\dots\\
& &\Psi^+\big(q_s-\frac{1}{2}\hbar k_s\big)\Psi\big(q_s+\frac{1}{2}\hbar 
k_s\big)\dots
\Psi\big(q_1+\frac{1}{2}\hbar k_s\big)\nonumber
\end{eqnarray}
allow us to consider them as some quasiprobabilities and provide useful bridge
between c- and q-cases:

\begin{equation}
<A>={\rm Tr}\rho A=\sum^{\infty}_{s=0}(s!)^{-1}\int\prod_{i=1}^s 
d\mu_iA_s(x_1,\dots,x_s)
W_s(x_1,\dots,x_s)
\end{equation}
The full description for quantum ensemble can be done by the whole hierarchy
of functions (symbols):

\begin{equation}
W=\{W_s(x_1,\dots,x_s), s=0,1,2\dots\}
\end{equation}
So, we may consider the following q-hierarchy as the result of ``wignerization'' 
procedure for c-BBGKY one:

\begin{eqnarray}
\partial_tW_s(t,x_1,\dots,x_s)&=&\sum^s_{j=1}L_j^0W_s(x_1,\dots,x_s)+
\sum_{j<n}\sum^s_{n=1}L_{j,n}W_s(x_1,\dots,x_s)\\
&+&
\sum^s_{j=1}\int 
dx_{s+1}\delta(k_{s+1})L_{j,s+1}W_{s+1}(x_1,\dots,x_{s+1})\nonumber
\end{eqnarray}
where

\begin{equation}
L^0_j=-\big(\frac{i}{m}\big)k_jp_j
\end{equation}

\begin{eqnarray}
L_{j,n}=(i\hbar)^{-1}\int d\ell \tilde{V_l}\Bigg[{\rm 
exp}\Bigg(-\frac{1}{2}\hbar\ell
\big(\frac{\partial}{\partial p_j}-\frac{\partial}{\partial p_n}\big)\Bigg)-
{\rm exp}\bigg(\frac{1}{2}\hbar\ell\big(\frac{\partial}{\partial p_j}-
\frac{\partial}{\partial p_n}\big)\Bigg)
\Bigg]{\rm exp}\Bigg(-\ell\big(\frac{\partial}{\partial 
k_j}-\frac{\partial}{\partial k_n}\big)\Bigg)
\end{eqnarray}

In quantum statistics the ensemble properties are described by the density
operator
\begin{equation}
\rho(t)=\sum_i w_i|\Psi_i(t)><\Psi_i(t)|, \quad \sum_iw_i=1
\end{equation}
After Weyl transform we have the following
 de\-com\-position via partial Wigner functions 
$W_i(p,q,t)$ for the whole ensemble Wigner function:
\begin{equation}
W(p,q,t)=\sum_iw_iW_i(p,q,t) 
\end{equation}
where the partial Wigner functions

\begin{eqnarray}
W_n(q,p,t)\equiv\frac{1}{2\pi\hbar}\int\ud\xi{\rm exp}\Big(-\frac{i}{\hbar}p\xi\Big)
\Psi^*_n(q-\frac{1}{2}\xi,t)\Psi_n(q+\frac{1}{2}\xi,t)
\end{eqnarray}
are solutions of proper Wigner equations:

\begin{eqnarray}
\frac{\partial W_n}{\partial t}=-\frac{p}{m}\frac{\partial W_n}{\partial q}+
\sum^{\infty}_{\ell=0}\frac{(-1)^\ell(\hbar/2)^{2\ell}}{(2\ell+1)!}
\frac{\partial^{2\ell+1}U_n(q)}{\partial q^{2\ell+1}}
\frac{\partial^{2\ell+1}W_n}{\partial p^{2\ell+1}}
\end{eqnarray}

Our approach, presented below, in some sense has allusion on the analysis of the following
standard simple model considered in [1].
Let us consider model of interaction of nonresonant atom with quantized electromagnetic field:
\begin{eqnarray}
\hat{H}=\frac{\hat{p}_x^2}{2m}+U(\hat{x}),\qquad
U(\hat{x})=U_0(z,t)g(\hat{x})\hat{a}^+\hat{a}
\end{eqnarray}
where potential $U$ 
depends on creation/annihilation operators and some polynomial on $\hat{x}$ 
operator function (or approximation)
$g(\hat{x})$.
It is possible to solve Schroedinger equation
\begin{eqnarray}
i\hbar\frac{\ud|\Psi>}{\ud t}=\hat{H}|\Psi>
\end{eqnarray}
by the simple ansatz 
\begin{eqnarray}
|\Psi(t)>=\sum_{-\infty}^{\infty}w_n\int\ud x \Psi_n(x,t)|x>\otimes|n>
\end{eqnarray}
which leads to the hierarchy of analogous equations with potentials created by 
n-particle Fock subspaces
\begin{eqnarray}
i\hbar\frac{\partial\Psi_n(x,t)}{\partial t}=\Big\{\frac{\hat{p}_x^2}{2m}+
 U_0(t)g(x)n\Big\}\Psi_n(x,t)
\end{eqnarray}
where
$\Psi_n(x,t)$ is the probability amplitude of finding the atom at 
the time $t$ at the position $x$ and the field in the $n$ Fock state.
Instead of this, we may apply the Wigner approach starting with proper full density matrix
\begin{eqnarray}
&&\hat{\rho}=|\Psi(t)><\Psi(t)|=\\
&&\sum_{n',n''}w_{n'}w^*_{n''}\int\ud x'\int\ud x''
\Psi_{n'}(x',t)\Psi^*_{n''}(x'',t)|x'><x''|\otimes|n'><n''|\nonumber
\end{eqnarray}
Standard reduction gives pure atomic density matrix
\begin{eqnarray}
&&\hat{\rho}_a\equiv\int^{\infty}_{n=0}<n|\hat{\rho}|n>=\\
&&\sum|w_n|^2
\int\ud x'\int\ud x''\Psi_n(x',t)\Psi^*_n(x'',t)|x'><x''|\nonumber
\end{eqnarray}
Then we have incoherent superposition 
\begin{equation}
W(x,p,t)=\sum^{\infty}_{n=0}|w_n|^2W_n(x,p,t)
\end{equation}
of the atomic Wigner functions (28)
corresponding to the atom motion in the potential $U_n(x)$ 
(which is not more than polynomial in $x$) generated by $n$-level Fock state. 
They are solutions of proper Wigner equations (29).

The next case describes the important decoherence process.
Let us have collective and environment subsystems with their own Hilbert spaces 
\begin{equation}
\mathcal{H}=\mathcal{H}_c\otimes\mathcal{H}_e
\end{equation}
Relevant dynamics is described by three parts including interaction
\begin{equation}
H=H_c\otimes I_e+I_c\otimes H_e+H_{int}
\end{equation}
For analysis, we can choose Lindblad master equation [1]

\begin{eqnarray}
\dot{\rho}=\frac{1}{i\hbar}[H,\rho]-
\sum_n\gamma_n(L^+_nL_n\rho+
\rho L^+_nL_n-2L_n\rho L^+_n)
\end{eqnarray}
which preserves the positivity of density matrix and it is Markovian
but it is not general form of exact master equation.
Other choice is Wigner transform of master equation and it is more preferable for us

\begin{eqnarray}
\dot{W}=\{H,W\}_{PB}+
\sum_{n\geq 1}\frac{\hbar^{2n}(-1)^n}{2^{2n}(2n+1)!}
\partial^{2n+1}_q U(q)\partial_p^{2n+1}W(q,p)+
2\gamma\partial_p pW+D\partial^2_pW
\end{eqnarray}

In the next section we consider the variational-wavelet approach for the solution of all
these Wigner-like equations (4), (5), (6), (23), (29), (40) for the case of an 
arbitrary polynomial $U(q, p)$, which corresponds to a finite number 
of terms in the series expansion in (5), (29), (40) 
or to proper finite order of $\hbar$. Analogous approach can be 
applied to classical counterpart (16) also.
Our approach is based on the extension of our variational-wavelet 
approach [2]-[8].
Wavelet analysis is some set of mathematical methods, which gives the possibility to
take into account high-localized states, control convergence of any type of expansions
and gives maximum sparse
forms for the general type of operators in such localized bases.
These bases are the natural generalization of standard coherent, 
squeezed, thermal squeezed states [1],
which correspond to quadratical systems (pure linear dynamics) with Gaussian Wigner functions.
The representations of underlying symmetry group (affine group in the simplest case) 
on the proper functional space of states generate the exact multiscale expansions
which allow to control contributions to
the final result from each scale of resolution from the whole underlying 
infinite scale of spaces. 
Numerical calculations according to methods of part I [9]
explicitly demonstrate the quantum interference of
generalized localized states, pattern formation from localized eigenmodes and 
the appearance of (stable) localized patterns (waveletons).

\section{VARIATIONAL MULTIRESOLUTION REPRESENTATION}

\subsection{Multiscale Decomposition for Space of States: 
Functional Realization and Metric Structure}

We obtain our multiscale/multiresolution representations for solutions of Wig\-ner-like equations
via a variational-wavelet approach. 
We represent the solutions as 
decomposition into localized eigenmodes (regarding action of affine group, i.e.
hidden symmetry of the underlying functional space of states) 
related to the hidden underlying set of scales [10]: 
\begin{eqnarray}
W_n(t,q,p)=\displaystyle\bigoplus^\infty_{i=i_c}W^i_n(t,q,p),
\end{eqnarray}
where value $i_c$ corresponds to the coarsest level of resolution
$c$ or to the internal scale with the number $c$ in the full multiresolution decomposition
of the underlying functional space ($L^2$, e.g.) corresponding to the problem under consideration:
\begin{equation}
V_c\subset V_{c+1}\subset V_{c+2}\subset\dots
\end{equation}
and $p=(p_1,p_2,\dots),\quad q=(q_1,q_2,\dots),\quad x_i=(p_1,q_1,\dots,p_i,q_i)$ 
are coordinates in phase space.
In the following we may consider as fixed as variable 
numbers of particles.

We introduce the Fock-like space structure (in addition to the standard one, 
if we consider second-quantized case) on the whole space of internal hidden scales.
\begin{eqnarray}
H=\bigoplus_i\bigotimes_n H^n_i
\end{eqnarray}
for the set of n-partial Wigner functions (states):
\begin{equation}
W^i=\{W^i_0,W^i_1(x_1;t),\dots,
W^i_N(x_1,\dots,x_N;t),\dots\},
\end{equation}
where
$W_p(x_1,\dots, x_p;t)\in H^p$,
$H^0=C,\quad H^p=L^2(R^{6p})$ (or any different proper functional spa\-ce), 
with the natural Fock space like norm: 
\begin{eqnarray}
(W,W)=W^2_0+
\sum_{i}\int W^2_i(x_1,\dots,x_i;t)\prod^i_{\ell=1}\mu_\ell.
\end{eqnarray}
First of all, we consider $W=W(t)$ as a function of time only,
$W\in L^2(R)$, via
multiresolution decomposition which naturally and efficiently introduces 
the infinite sequence of the underlying hidden scales [10].
We have the contribution to
the final result from each scale of resolution from the whole
infinite scale of spaces (16).
The closed subspace
$V_j (j\in {\bf Z})$ corresponds to  the level $j$ of resolution, 
or to the scale j
and satisfies
the following properties:
let $D_j$ be the orthonormal complement of $V_j$ with respect to $V_{j+1}$: 
$
V_{j+1}=V_j\bigoplus D_j.
$
Then we have the following decomposition:
\begin{eqnarray}
\{W(t)\}=\bigoplus_{-\infty<j<\infty} D_j 
=\overline{V_c\displaystyle\bigoplus^\infty_{j=0} D_j},
\end{eqnarray}
in case when $V_c$ is the coarsest scale of resolution.
The subgroup of translations generates a basis for the fixed scale number:
$
{\rm span}_{k\in Z}\{2^{j/2}\Psi(2^jt-k)\}=D_j.
$
The whole basis is generated by action of the full affine group:
\begin{eqnarray}
{\rm span}_{k\in Z, j\in Z}\{2^{j/2}\Psi(2^jt-k)\}=
{\rm span}_{k,j\in Z}\{\Psi_{j,k}\}
=\{W(t)\}
\end{eqnarray}

\subsection{Tensor Product Structure}

Let sequence 
\begin{eqnarray}
\{V_j^t\},\qquad V_j^t\subset L^2(R)
\end{eqnarray}
correspond to multiresolution analysis on time axis and 
\begin{equation}
\{V_j^{x_i}\},\qquad V_j^{x_i}\subset L^2(R)
\end{equation}
correspond to multiresolution analysis for coordinate $x_i$,
then
\begin{equation}
V_j^{n+1}=V^{x_1}_j\otimes\dots\otimes V^{x_n}_j\otimes  V^t_j
\end{equation}
corresponds to multiresolution analysis for n-particle distribution fuction 
$W_n(x_1,\dots,x_n;t)$.
E.g., for $n=2$:
\begin{eqnarray}
V^2_0=\{f:f(x_1,x_2)=
\sum_{k_1,k_2}a_{k_1,k_2}\phi^2(x_1-k_1,x_2-k_2),\qquad
a_{k_1,k_2}\in\ell^2(Z^2)\},
\end{eqnarray}
where 
\begin{equation}
\phi^2(x_1,x_2)=\phi^1(x_1)\phi^2(x_2)=\phi^1\otimes\phi^2(x_1,x_2),
\end{equation}
and $\phi^i(x_i)\equiv\phi(x_i)$ form a multiresolution basis corresponding to
$\{V_j^{x_i}\}$.
If 
\begin{equation}
\{\phi^1(x_1-\ell)\},\ \ell\in Z
\end{equation}
 form an orthonormal set, then 
\begin{equation}
\phi^2(x_1-k_1, x_2-k_2)
\end{equation}
 form an orthonormal basis for $V^2_0$.
Action of affine group provides us by multiresolution representation of
$L^2(R^2)$. After introducing detail spaces $D^2_j$, we have, e.g. 
\begin{equation}
V^2_1=V^2_0\oplus D^2_0.
\end{equation}
Then
3-component basis for $D^2_0$ is generated by translations of three functions 
\begin{eqnarray}
&&\Psi^2_1=\phi^1(x_1)\otimes\Psi^2(x_2),\nonumber\\
&&\Psi^2_2=\Psi^1(x_1)\otimes\phi^2(x_2), \\
&&\Psi^2_3=\Psi^1(x_1)\otimes\Psi^2(x_2)\nonumber
\end{eqnarray}

Also, we may use the rectangle lattice of scales and one-dimentional wavelet
decomposition :
\begin{equation}
f(x_1,x_2)=\sum_{i,\ell;j,k}<f,\Psi_{i,\ell}\otimes\Psi_{j,k}>
\Psi_{j,\ell}\otimes\Psi_{j,k}(x_1,x_2),
\end{equation}
where bases functions
\begin{equation} 
\Psi_{i,\ell}\otimes\Psi_{j,k}
\end{equation}
depend on
two scales $2^{-i}$ and $2^{-j}$.

After construction the multidimensional bases  
we obtain our multiscale\-/mul\-ti\-re\-so\-lu\-ti\-on 
representations for observables (operators, symbols), states, partitions
via the variational approaches [2]-[8] as for c-BBGKY as for its quantum counterpart
and related reductions but before we need to construct reasonable multiscale 
decomposition for all operators included in the set-up.

\subsection{FWT Decomposition for Observables}

One of the key point of wavelet analysis approach, 
the so called Fast Wavelet Transform (FWT), 
demonstrates that for a large class of
operators the wavelet functions are good approximation for true eigenvectors; and the corresponding 
matrices are almost diagonal. FWT gives  the maximum sparse form for wide classes 
of operators [10].
So, let us denote our (integral/differential) operator from equations under 
consideration  
as  $T$ ($L^2(R^n)\rightarrow L^2(R^n)$) and its kernel as $K$.
We have the following representation:
\begin{equation}
<Tf,g>=\int\int K(x,y)f(y)g(x)\ud x\ud y.
\end{equation}
In case when $f$ and $g$ are wavelets $\varphi_{j,k}=2^{j/2}\varphi(2^jx-k)$, 
(21) provides the standard representation for operator $T$.
Let us consider multiresolution representation
$
\dots\subset V_2\subset V_1\subset V_0\subset V_{-1}
\subset V_{-2}\dots
$. 
The basis in each $V_j$ is 
$\varphi_{j,k}(x)$,
where indices $\ k, j$ represent translations and scaling 
respectively. 
Let $P_j: L^2(R^n)\rightarrow V_j$ $(j\in Z)$ be projection
operators on the subspace $V_j$ corresponding to level $j$ of resolution:
$
(P_jf)(x)=\sum_k<f,\varphi_{j,k}>\varphi_{j,k}(x).
$ 
Let
$Q_j=P_{j-1}-P_j$ be the projection operator on the subspace $D_j$ ($V_{j-1}=V_j\oplus D_j$), 
then
we have the following 
representation of operator T which takes into account contributions from
each level of resolution from different scales starting with the
coarsest and ending to the finest scales [10]:
\begin{equation}
T=\sum_{j\in Z}(Q_jTQ_j+Q_jTP_j+P_jTQ_j).
\end{equation}
We need to remember that this is a result of presence of affine group inside this
construction.
The non-standard form of operator representation is a representation of
operator T as  a chain of triples
$T=\{A_j,B_j,\Gamma_j\}_{j\in Z}$, acting on the subspaces $V_j$ and
$D_j$:
$
 A_j: D_j\rightarrow D_j, B_j: V_j\rightarrow D_j,
\Gamma_j: D_j\rightarrow V_j,
$
where operators $\{A_j,B_j,\Gamma_j\}_{j\in Z}$ are defined
as
$A_j=Q_jTQ_j, \quad B_j=Q_jTP_j, \quad\Gamma_j=P_jTQ_j.$
The operator $T$ admits a recursive definition via
\begin{eqnarray}
T_j=
\left(\begin{array}{cc}
A_{j+1} & B_{j+1}\\
\Gamma_{j+1} & T_{j+1}
\end{array}\right),
\end{eqnarray}
where $T_j=P_jTP_j$ and $T_j$ acts on $ V_j: V_j\rightarrow V_j$.
So, it is possible to provide the following ``sparse'' action of operator $T_j$
on elements $f$ of functional realization of our space of states $H$:
\begin{equation}
(T_j f)(x)=\sum_{k\in Z}\left(2^{-j}\sum_{\ell}r_\ell f_{j,k-\ell}\right)
\varphi_{j,k}(x),
\end{equation}
in the wavelet basis $\varphi_{j,k}(x)=2^{-j/2}\varphi(2^{-j}x-k)$, where
\begin{equation}
f_{j,k-1}=2^{-j/2}\int f(x)\varphi(2^{-j}x-k+\ell)\ud x
\end{equation}
are wavelet coefficients and $r_\ell$  
are the roots of some additional linear system of equations related to
the ``type of localization'' [10].
So, we have the simple linear para\-met\-rization of
matrix representation of  our operators in localized wavelet bases
and of the action of
this operator on arbitrary vector/state in proper functional space.

\subsection{Variational Approach}

Now, after preliminary work with (functional) spaces, states and operators, 
we may apply our variational approach from [2]-[8].

Let $L$ be an arbitrary (non)li\-ne\-ar dif\-fe\-ren\-ti\-al\-/\-in\-teg\-ral operator 
 with matrix dimension $d$
(finite or infinite), 
which acts on some set of functions
from $L^2(\Omega^{\otimes^n})$:  
$\quad\Psi\equiv\Psi(t,x_1,x_2,\dots)=\Big(\Psi^1(t,x_1,x_2,\dots), \dots$,
$\Psi^d(t,x_1,x_2,\dots)\Big)$,
 $\quad x_i\in\Omega\subset{\bf R}^6$, $n$ is the number of particles:
\begin{eqnarray}
L\Psi&\equiv& L(Q,t,x_i)\Psi(t,x_i)=0,\\
Q&\equiv& Q_{d_0,d_1,d_2,\dots}(t,x_1,x_2,\dots,
\partial /\partial t,\partial /\partial x_1,
\partial /\partial x_2,\dots,
\int \mu_k)\nonumber\\
&=&
\sum_{i_0,i_1,i_2,\dots=1}^{d_0,d_1,d_2,\dots}
q_{i_0i_1i_2\dots}(t,x_1,x_2,\dots)
\Big(\frac{\partial}{\partial t}\Big)^{i_0}\Big(\frac{\partial}{\partial x_1}\Big)^{i_1}
\Big(\frac{\partial}{\partial x_2}\Big)^{i_2}\dots\int\mu_k\nonumber 
\end{eqnarray}
Let us consider now the $N$ mode approximation for the solution as 
the following ansatz:
\begin{eqnarray}
\Psi^N(t,x_1,x_2,\dots)=
\sum^N_{i_0,i_1,i_2,\dots=1}a_{i_0i_1i_2\dots}
 A_{i_0}\otimes 
B_{i_1}\otimes C_{i_2}\dots(t,x_1,x_2,\dots)
\end{eqnarray}
We shall determine the expansion coefficients from the following conditions
(related to proper choosing of variational approach):
\begin{eqnarray}
&&\ell^N_{k_0,k_1,k_2,\dots}\equiv 
\int(L\Psi^N)A_{k_0}(t)B_{k_1}(x_1)C_{k_2}(x_2)\ud t\ud x_1\ud x_2\dots=0
\end{eqnarray}
Thus, we have exactly $dN^n$ algebraical equations for  $dN^n$ unknowns 
$a_{i_0,i_1,\dots}$.
This variational ap\-proach reduces the initial problem 
to the problem of solution 
of functional equations at the first stage and 
some algebraical problems at the second one.
It allows to unify the multiresolution expansion with variational 
construction[2]-[8]. 

As a result, the solution is parametrized by the solutions of two sets of 
reduced algebraical
problems, one is linear or nonlinear
(depending on the structure of the generic operator $L$) and the rest are linear
problems related to the computation of the coefficients of reduced 
algebraic equations. It is also related to the choice of exact measure of localization
(including class of smothness) which are proper for our set-up.
These coefficients can be found  by some functional/algebraic methods
by using the
compactly supported wavelet basis functions or any other wavelet families [10].

As a result the solution of the equations/hierarchies from Section 1, as in c-
as in q-region, has the 
following mul\-ti\-sca\-le or mul\-ti\-re\-so\-lu\-ti\-on decomposition via 
nonlinear high\--lo\-ca\-li\-zed eigenmodes 
{\setlength\arraycolsep{0pt}
\begin{eqnarray}
&&W(t,x_1,x_2,\dots)=
\sum_{(i,j)\in Z^2}a_{ij}U^i\otimes V^j(t,x_1,\dots),\nonumber\\
&&V^j(t)=
V_N^{j,slow}(t)+\sum_{l\geq N}V^j_l(\omega_lt), \ \omega_l\sim 2^l, 
\end{eqnarray}
}
$$U^i(x_s)=
U_M^{i,slow}(x_s)+\sum_{m\geq M}U^i_m(k^{s}_mx_s), \ k^{s}_m\sim 2^m,
$$
which corresponds to the full multiresolution expansion in all underlying time/space 
scales.
The formulae (67) give the expansion into a slow part
and fast oscillating parts for arbitrary $N, M$.  So, we may move
from the coarse scales of resolution to the 
finest ones for obtaining more detailed information about the dynamical process.
In this way one obtains contributions to the full solution
from each scale of resolution or each time/space scale or from each nonlinear eigenmode.
It should be noted that such representations 
give the best possible localization
properties in the corresponding (phase)space/time coordinates. 
Formulae (67) do not use perturbation
techniques or linearization procedures.
Numerical calculations are based on compactly supported
wavelets and wavelet packets and on evaluation of the 
accuracy on 
the level $N$ of the corresponding cut-off of the full system 
regarding norm (45):
\begin{equation}
\|W^{N+1}-W^{N}\|\leq\varepsilon.
\end{equation}

\section{MODELING OF PATTERNS}

To summarize, the key points are:

1. The ansatz-oriented choice of the (multi\-di\-men\-si\-o\-nal) ba\-ses
related to some po\-ly\-no\-mi\-al tensor algebra. 

2. The choice of proper variational principle. A few 
pro\-je\-c\-ti\-on/ \-Ga\-ler\-kin\--li\-ke 
principles for constructing (weak) solutions are considered.
The advantages of formulations related to biorthogonal
(wavelet) decomposition should be noted. 

3. The choice of  bases functions in the scale spaces $D_j$ from wavelet zoo. They 
correspond to high-localized (nonlinear) oscillations/excitations, 
nontrivial local (stable) distributions/fluctuations,
etc. Besides fast convergence properties it should be noted 
minimal complexity of all underlying calculations, especially in case of choice of wavelet
packets which minimize Shannon entropy. 

4.  Operator  representations providing maximum sparse representations 
for arbitrary (pseudo) differential/ integral operators 
$\ud f/\ud x$, $\ud^n f/\ud x^n$, $\int T(x,y)f(y)\ud y)$, etc.

5. (Multi)linearization. Besides the variation approach we can consider also a different method
to deal with (polynomial) nonlinearities: para-products-like decompositions.

To classify the qualitative behaviour we apply
standard methods from general control theory or really use the 
control.
We will start from a priori unknown coefficients, the exact values of which 
will subsequently be recovered.
Roughly speaking, we will fix only class of nonlinearity 
(polynomial in our case)
which covers a broad variety of examples of possible truncation of the systems.
As a simple model we choose band-triangular 
non-sparse matrices $(a_{ij})$.
These matrices provide tensor structure of bases in (extended) phase space
and are generated by the roots of the reduced variational (Galerkin-like) 
systems.
As a second step we need to restore the coefficients from these 
matrices
by which we may classify the types of behaviour. 
We start with the localized mode, which is a base mode/eigenfunction,
(Fig. 1, 9, 10 from Part I), corresponding to definitions from Section 2.2, Part I, 
which was constructed as a tensor product of the two Daubechies functions. 
Fig.~1, 4 below demonstrate 
the result of summation of series (67) up to value of the 
dilation/scale parameter equal to four and six, respectively.
It's done in the bases of symmlets [10] with the corresponding matrix 
elements equal to one.  The size of matrix of ``Fourier-wavelet coefficients''
is 512x512. So, different possible  distributions of  the root values 
of the generical 
algebraical systems (66) provide qualitatively different types of behaviour.
Generic algebraic system (66), Generalized Dispersion Relation (GDR), 
provide the possibility for algebraic control. 
The above choice
provides us by a distribution with chaotic-like equidistribution.
But, if we consider a band-like structure of matrix $(a_{ij})$ 
with the band along the main diagonal with 
finite size ($\ll 512$) and values, e.g. five, while the other 
values are equal to one, we obtain
localization in a fixed finite area of the full phase space, 
i.e. almost all energy of the system 
is concentrated in this small volume. This corresponds to waveleton states and 
is shown in Fig.~2, constructed by means of Daubechies-based wavelet packets. 
Depending on the type of 
solution,  such localization may be conserved during the whole time evolution
(asymptotically-stable) or up to the needed value from the whole time scale (e.g. enough 
for plasma fusion/confinement in the case of fusion modeling by means of c-BBGKY hierarchy
for dynamics of partitions).

\section{CONCLUSION}

So, by using wavelet bases with their best (phase) space/time      
localization  properties we can describe the localized (coherent) structures in      
quantum systems with complicated behaviour (Fig.~1, 4).
The modeling demonstrates the formation of different (stable) pattern or orbits 
generated by internal hidden symmetry from
high-localized structures.
Our (nonlinear) eigenmodes are more realistic for the modelling of 
nonlinear classical/quantum dynamical process  than the corresponding linear gaussian-like
coherent states. Here we mention only the best convergence properties of the expansions 
based on wavelet packets, which  realize the minimal Shannon entropy property
and the exponential control of convergence of expansions like (67) based on the norm (45).
Fig.~2 corresponds to (possible) result of superselection
(einselection) [1] after decoherence process started from entangled state (Fig.~5);
Fig.~3 and Fig.~6 demonstrate the steps of multiscale resolution during modeling  
of entangled states leading to the growth of degree of entanglement.
It should be noted that
we can control the type of behaviour on the level of the reduced algebraical variational 
system, GDR (66).

\twocolumn
\newpage

\begin{figure}
\begin{center}
\begin{tabular}{c}
\includegraphics*[width=60mm]{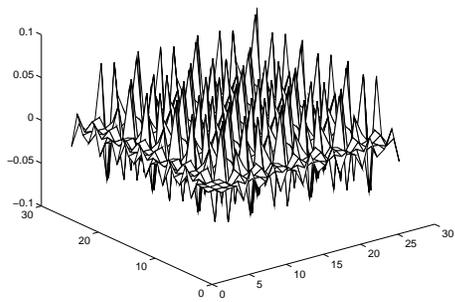}
\end{tabular}
\end{center}
\caption{Level 4 MRA.}
\end{figure}

\begin{figure}
\begin{center}
\begin{tabular}{c}
\includegraphics*[width=80mm]{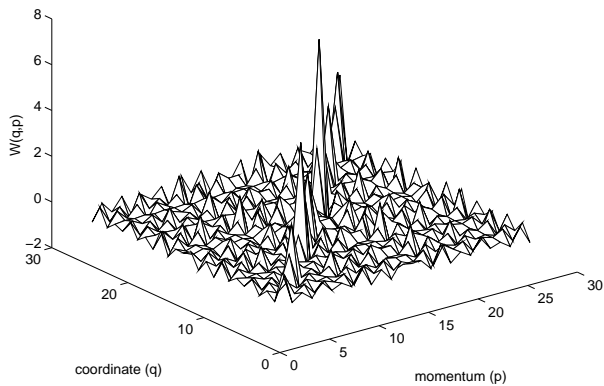}
\end{tabular}
\end{center}
\caption{Localized pattern, (waveleton) Wigner function.}
\end{figure}

\begin{figure}
\begin{center}
\begin{tabular}{c}
\includegraphics*[width=60mm]{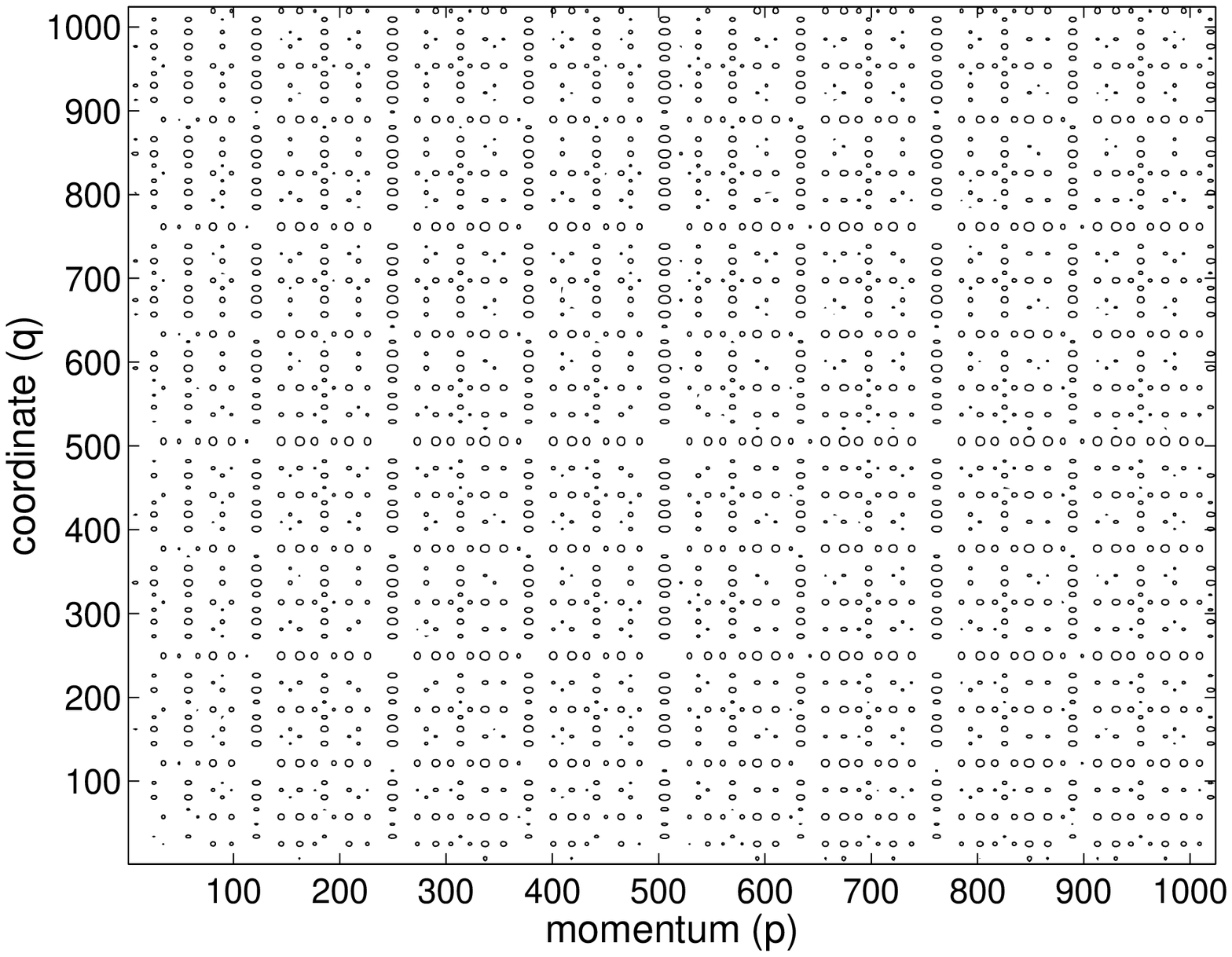}
\end{tabular}
\end{center}
\caption{Interference picture on the level 4 approximation for Wigner function.}
\end{figure}

\begin{figure}
\begin{center}
\begin{tabular}{c}
\includegraphics*[width=60mm]{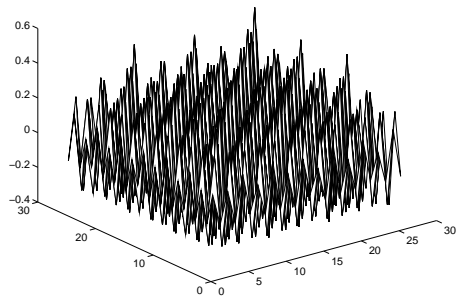}
\end{tabular}
\end{center}
\caption{Level 6 MRA.}
\end{figure}

\begin{figure}
\begin{center}
\begin{tabular}{c}
\includegraphics*[width=80mm]{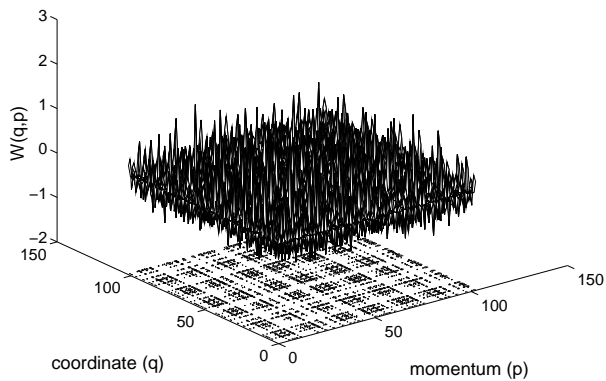}
\end{tabular}
\end{center}
\caption{Entangled-like Wigner function.}
\end{figure}

\begin{figure}
\begin{center}
\begin{tabular}{c}
\includegraphics*[width=60mm]{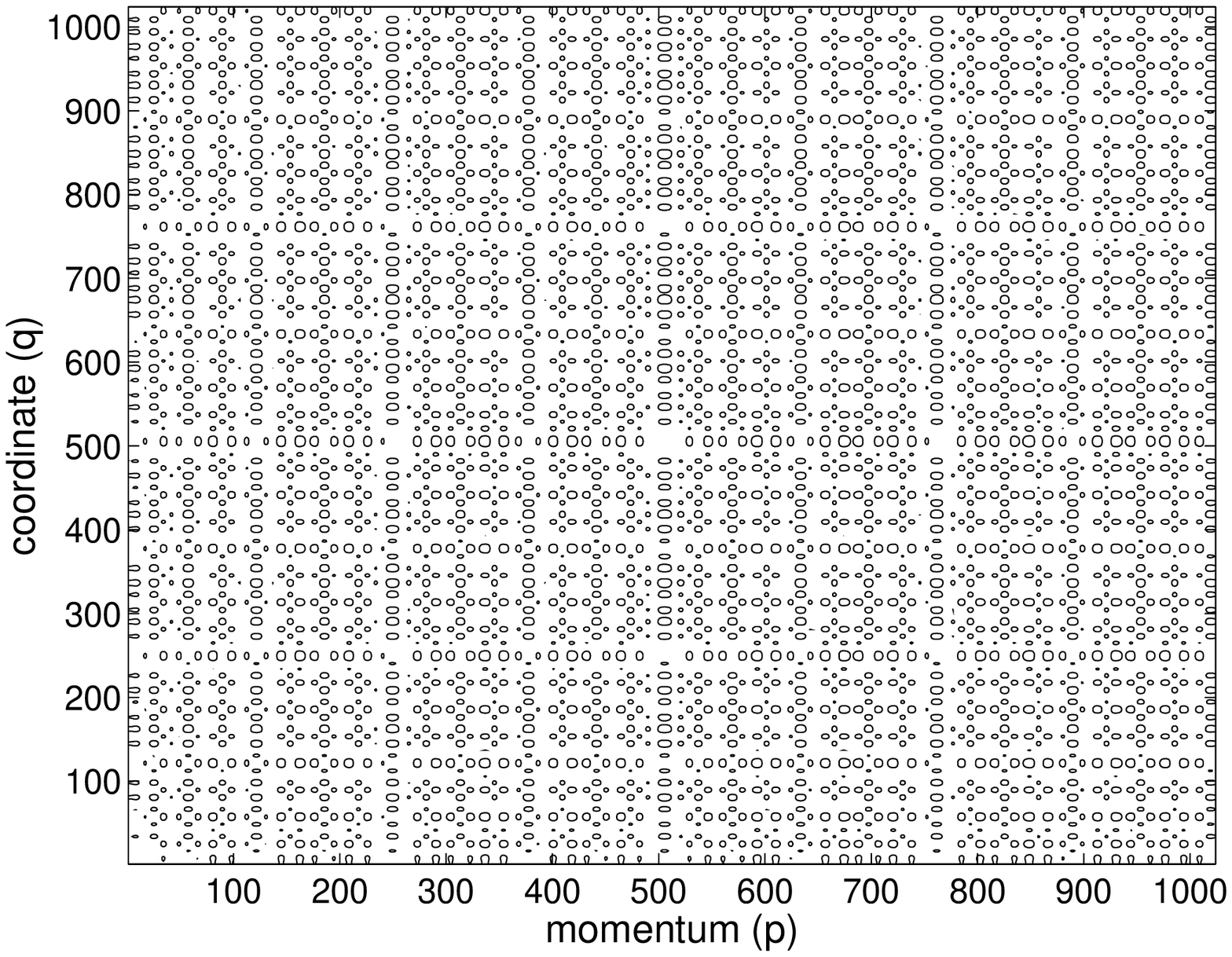}
\end{tabular}
\end{center}
\caption{Interference picture on the level 6 approximation for Wigner function.}
\end{figure}

\onecolumn

\end{document}